\documentclass{emulateapj}
\begin{document}

\title{The XMM Cluster Survey: A Massive Galaxy Cluster at $z=1.45$}

\author{S.A.\ Stanford\altaffilmark{1,}\altaffilmark{2}}

\author{A.\ Kathy Romer\altaffilmark{3}}

\author{Kivanc Sabirli\altaffilmark{4}}

\author{Michael Davidson\altaffilmark{5}}

\author{Matt Hilton\altaffilmark{6}}

\author{Pedro T.P.\ Viana\altaffilmark{7,}\altaffilmark{8}}

\author{Chris A.\ Collins\altaffilmark{6}}

\author{Scott T.\ Kay\altaffilmark{9}}

\author{Andrew R.\ Liddle\altaffilmark{3}}

\author{Robert G.\ Mann\altaffilmark{5}}

\author{Christopher J.\ Miller\altaffilmark{10}}

\author{Robert C.\ Nichol\altaffilmark{11}}

\author{Michael J.\ West\altaffilmark{12,}\altaffilmark{13}} 

\author{Christopher J.\ Conselice\altaffilmark{14}}

\author{Hyron Spinrad\altaffilmark{15}}

\author{Daniel Stern\altaffilmark{16}}

\author{Kevin Bundy\altaffilmark{17}}

\altaffiltext{1}{University of California, Davis, CA 95616; adam@igpp.ucllnl.org}

\altaffiltext{2}{Institute of
Geophysics and Planetary Physics, Lawrence Livermore National
Laboratory, Livermore, CA 94551} 

\altaffiltext{3}{Astronomy Centre, University of Sussex, Falmer, Brighton, BN1 9QJ, U.K.}

\altaffiltext{4}{Department of Physics, Carnegie Mellon University, 5000 Forbes Ave., Pittsburgh, PA, 15217}

\altaffiltext{5}{Institute of Astronomy, University of Edinburgh, Blackford Hill, Edinburgh, EH9 9HJ, U.K.}

\altaffiltext{6}{Astrophysics Research Institute, Liverpool John Moores University, Twelve Quays House, Egerton Wharf, Birkinhead CH41 1LD, U.K.}

\altaffiltext{7}{Departamento de Matema\'tica Aplicada da Faculdade de Ciencias da Universidade do Porto, Rua do Campo Alegre, 687, 4169-007, Portugal}

\altaffiltext{8}{Centro de Astrofisca da Universidade do Porto, Rua das Estrelas, 4150-762, Porto, Portugal}

\altaffiltext{9}{Astrophysics, Keble Road, Oxford, OX1 3RH, U.K.}

\altaffiltext{10}{Cerro-Tololo Inter-American Observatory, National Optical Astronomy Observatory, 950 North Cherry, Tucson, AZ, 85719}

\altaffiltext{11}{Institute of Cosmology and Gravitation, University of Portsmouth, Portsmouth, PO1 2EG, U.K.}

\altaffiltext{12}{Department of Physics and Astronomy,
University of Hawaii, 200 W. Kawili Street, Hilo, Hawaii 96720,
U.~S.~A.}

\altaffiltext{13}{Gemini Observatory, Casilla 603, La Serena, Chile}

\altaffiltext{14}{University of Nottingham, University Park, Nottingham, NG9 2RD, UK}

\altaffiltext{15}{Astronomy Department, University of California, Berkeley, CA 94720}

\altaffiltext{16}{Jet Propulsion Laboratory, California Institute of Technology, MS 169-506, 4800 Oak Grove Road., Pasadena, CA, 91109}

\altaffiltext{17}{California Institute of Technology, MS 105-24, 1200 California Blvd., Pasadena, CA, 91125}
 
\newpage

\shorttitle{MASSIVE GALAXY CLUSTER AT $z=1.45$}
\shortauthors{STANFORD ET AL.}

\begin{abstract} 

We report the discovery of XMMXCS~J2215.9-1738, a massive galaxy
cluster at $z =1.45$, which was found in the {\em XMM Cluster Survey}.
The cluster candidate was initially identified as an extended X-ray
source in archival XMM data.  Optical spectroscopy shows that 6
galaxies within a $\sim$60 arcsec diameter region lie at $z = 1.45\pm
0.01$.  Model fits to the X-ray spectra of the extended emission yield
$kT = 7.4^{+2.7}_{-1.8}$ keV (90\% confidence); if there is an
undetected central X-ray point source then $kT = 6.5^{+2.6}_{-1.8}$
keV.  The bolometric X-ray luminosity is $L_x = 4.4^{+0.8}_{-0.6}
\times 10^{44}$ ergs s$^{-1}$ over a 2 Mpc radial region.  The
measured $T_x$, which is the highest known for a cluster at $z > 1$,
suggests that this cluster is relatively massive for such a high
redshift. The redshift of XMMXCS~J2215.9-1738 is the highest currently
known for a spectroscopically-confirmed cluster of galaxies.
    
\end{abstract}


\section{Introduction}

High redshift galaxy clusters provide important laboratories for the
study of structure formation and galaxy evolution. They also can be
used to constrain cosmological parameters independent of the Cosmic
Microwave Background and supernova methods. The number of known
clusters at $z>1$ is currently small ($\simeq 10$ with X-ray
confirmation: e.g. \citealt{lynx,0910,1252,mullis}), but growing rapidly
thanks to new surveys being carried out in the X-ray
\citep*[e.g.][]{xcs,mullis,xmmlss,champ}, optical \citep*[e.g.][]{RCS}, 
and infrared \citep*[e.g.][]{bootes}. In the
near future, Sunyaev-Zel'dovich surveys should also provide large
numbers of high redshift clusters \citep{sz}.  We announce here the
discovery of a high redshift cluster in the XMM Cluster Survey 
\citep*[XCS:][]{xcs} and the measurement of its X-ray temperature. At
$z=1.45$, XMMXCS~J2215.9-1738 is the most distant cluster
with spectroscopic redshift confirmation to date.  
Unless otherwise noted, we assume $\Omega_m = 0.27$, $\Lambda = 0.73$, 
and $H_o = 70$~km s$^{-1}$ Mpc$^{-1}$.

\section{Observations}

\subsection{X-ray}

XMM-Newton is the most sensitive X-ray spectral-imaging telescope
deployed to date.  The XCS identifies cluster candidates in the {\it
XMM} data archive using the signature of X-ray extent. With a
projected total area of 500 deg$^2$, the XCS is expected to catalog
several thousand clusters out to a redshift of $z\simeq2$
\citep{xcs}. To date, 1,847 {\it XMM} observations have been
processed using an automated pipeline, resulting in a survey area of
168 deg$^2$ at $b>20^\circ$. This is the net area available for
cluster searching and excludes the Magellanic Clouds, artifacts due to
out of time events, and X-ray sources that extend over a large
fraction of the field of view, such as supernova remnants, Galactic
clusters, or low redshift galaxy clusters.  The initial catalog
contains 1764 cluster candidates (see Davidson et al.\ in preparation
and Sabilri et al.\ in preparation for details). Here we focus on 5
moderately deep observations of varying length in the direction of the
$z=2.217$ quasar LBQS 2212-1759, which were obtained in 2000 and
2001. An extended source was detected at the 4 $\sigma$ or greater
level by the XCS pipeline at the same location, $9'$ from the center
of the field of view, in 3 of the 5 observations.  The combined {\it
XMM} exposure time, after correction for solar flares etc., in the
direction of the cluster candidate is 237 ks (MOS1).  The typical
point source positional uncertainty in the XCS is 3 arcsec at 9 arcmin
off-axis (Davidson et al. in preparation) and we have verified this by
checking the positions of three quasars in the summed X-ray data near
the cluster candidate.

\subsection{Optical and IR Imaging}

An $I$-band image of the area around XMMXCS~J2215.9-1738 was
found in the ESO Imaging Survey \citep{eis}.  This 9000~s image reaches a limiting magnitude of
$24$ (Vega), and allows even $z > 1$ early-type galaxies to be
detected.  On the basis of visual inspection of the $I$-band image,
XMMXCS~J2215.9-1738 was selected as a possible $z > 1$ galaxy cluster, worthy
of spectroscopic followup.

\begin{figure}[h]
\epsscale{1.1}
\plotone{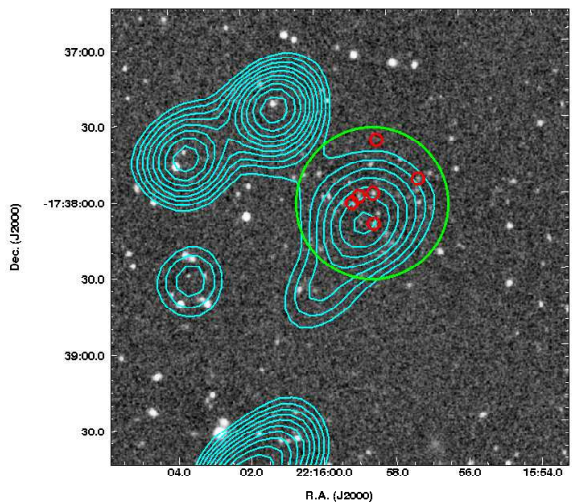}

\caption{$K_s$-band image of a 3 arcmin field
centered on the cluster with an X-ray overlay of contours in blue.
The 6 spectroscopically confirmed members are circled in red (ID$=$14651 is
not detected in the $K_s$ image). The green circle of diameter 60 arcsec shows 
the location of the region used for the X-ray analysis reported in section~\ref{secxspec}.}

\label{kimg}
\end{figure}

Subsequent to the first slit mask spectroscopy (see below), $K_s$-band
imaging was carried out on the 200~inch telescope at Palomar on 2005 
October 21.  A total of 3240~s of exposure time was obtained in clear
conditions, and the images were reduced following standard infrared
procedures.  The reduced image was calibrated onto the Vega system by
comparison of instrumental magnitudes for stars with the same objects
in the 2MASS database.  The image is shown in Figure~\ref{kimg}, along
with contours of the X-ray emission.  The X-ray emission from the
cluster candidate is clearly extended compared to the three nearby
point sources.  None of the X-ray sources in Figure~\ref{kimg} (the cluster,
the three point sources, or the southern extension) has been cataloged
previously, based on an examination of the NED and SIMBAD databases.

\begin{figure}[h]
\epsscale{0.9}
\plotone{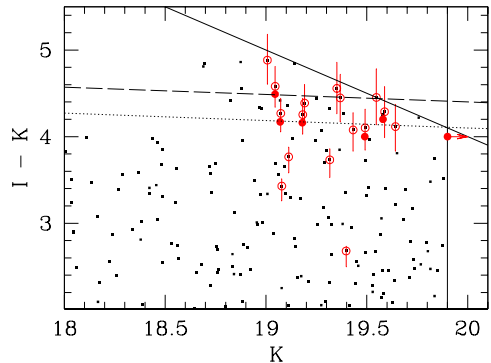}

\caption{An $I-K_s$ vs $K_s$ color-magnitude diagram for XMMXCS~J2215.9-1738; 
aperture magnitudes are used for the colors and $K_s$ magnitudes.  The
black points are all the objects within a 3 arcmin box around the
position of the X-ray source, and the objects circled in red lie
within 30 arcsec of the center of the X-ray source.  The solid red
points are those with concurrent spectroscopic redshifts at $z=1.45$.
The dashed (dotted) line represents the expected color-magnitude
relation for galaxies as predicted by a Bruzual \& Charlot model in
which all the stars formed within a 0.1 Gyr burst starting at $z_f =
3$ ($z_f = 2$) in the standard cosmology.  The solid lines represent
the 5 $\sigma$ limits to the $K_s$ and $I - K_s$ photometry. }

\label{cmd}
\end{figure}

The $I$ and $K_s$ band images were photometered using SExtractor.
Objects were detected on the $K_s$ image and then aperture magnitudes
of diameter $2.5$ arcsec were measured on both images using the same
SExtractor catalog.  The resulting color-magnitude diagram shown in
Figure~\ref{cmd} was used for target selection for the second mask in
the spectroscopic followup program.

\subsection{Keck Spectroscopy}

The first spectroscopic followup was carried out using the Deep
Imaging Multi-Object Spectrograph (DEIMOS: Faber et al.\ 2003) at the
Keck II telescope on UT 2005 September 1.  Targets were selected from
the $I$-band image to have $23 < I < 24$ and be located in the central
few arcmin of the area around the X-ray candidate cluster.  The slits had
widths of 1.0 arcsec and minimum lengths of 5 arcsec.  We used the
600ZD grating, which is blazed at 7500 \AA, to cover a nominal
wavelength range of 5000 to 10000 \AA.  The dispersion of $\sim$0.65 \AA~pixel$^{-1}$
resulted in a spectral resolution of $3.8$\AA.  Of particular note is
that this spectral resolution is adequate for being able to detect
both lines in the [O II]$\lambda,3727$ doublet.  We obtained 5
$\times$ 1800~s exposures with this setup in photometric conditions
with 1.1 arcsec seeing.  The observations were carried out with the
slitlets aligned close to the parallactic angle.

A second slit mask was observed, again using DEIMOS at the Keck II
telescope, on UT 2005 November 5.  For this mask, we were able to use
the color information seen in Figure~\ref{cmd} to select targets
likely to be in the cluster by choosing objects in the area of the
color-magnitude diagram where the red sequence is expected for a
$z=1.45$ cluster.  These observations were conducted in poor
conditions, with $\sim$1.5 arcsec seeing.  A total of 5400~s of
exposure time was obtained with the second mask.
  
The slitmask data for the two masks were separately reduced using the
DEEP pipeline.  A relative flux calibration was obtained from
longslit observations of the standard stars G191B2B and Wolf 1346.
One--dimensional spectra were extracted from the sum of the
reduced data for each slitlet.

\section{Results}

\subsection{X-ray Analysis}
\label{secxspec}

\begin{figure}[h]

\plotone{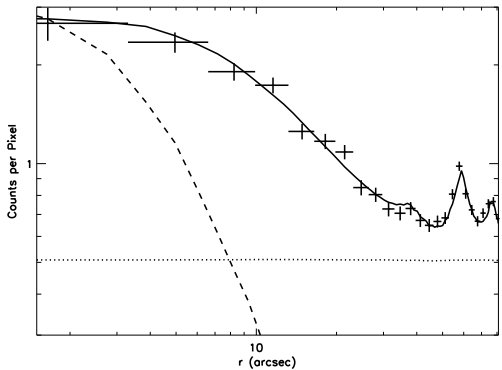}

\caption{The spatial profile of the X-ray emission.  The points are the data.  
The best fitting model, combining both the extended component and the
point sources (which are obvious as bumps 
at large radii), is shown by
the solid line.  A PSF is shown by the dashed line.  The background
is shown by the horizontal dotted line.}
\label{xprof}
\end{figure}

To examine the X-ray emission from the intracluster medium, we need to
account for contamination from the nearby point sources.  At the same
time it behooves us to allow for a possible central point source, such
as a galaxy with an active galactic nucleus, that would affect our
attempts to extract information on e.g.\ the temperature of the
ICM.  Using SHERPA, we carried out 2-dimensional fitting of the detected nearby
point sources simultaneously with fitting of the extended emission
from the cluster candidate.  In addition to the obvious point
sources to the east of the cluster, we have included a variable
strength point source fixed at the center of the extended emission.
The profile of the X-ray emission is shown in Figure~\ref{xprof}.  The
best fitting $\beta$ model combined with the point sources is
shown, along with a PSF which indicates that extended emission is
present. The brightest point sources are
obvious as bumps in the profile at radii of $\sim$37, 60, and 75 arcsec.
We find that a central point source contributes at most 10\% of the
counts within the central $r = 30$ arcsec region.  The best fitting
$\beta$ model for the extended component has a core radius of $r_c =
73^{+18}_{-16}$ kpc and $\beta = 0.53^{+0.04}_{-0.03}$.

\begin{figure}[h]
\includegraphics[angle=270,width=0.5\textwidth]{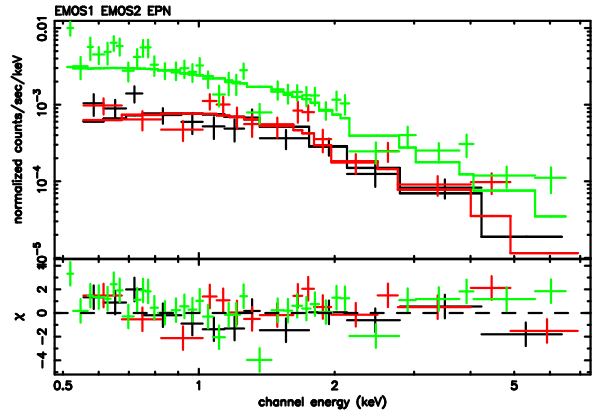}

%\epsscale{0.9}
\caption{X-ray spectrum of XMMXCS~J2215.9-1738. 
Counts from EMOS1, EMOS2, and EPN are plotted with black, red, and
green respectively.  The spectra are after accounting for the point sources, 
including a possible central source.  The best fit to these spectra is an absorbed MEKAL
model with properties specified in the text; the folded model is
overplotted (only $T_x$ and the normalisation are left as free
parameters).}
\label{xspec}
\end{figure}

Next we used the results of this two-dimensional fitting procedure to
accurately determine the spectrum of the extended emission after
correction for contamination by the point sources.  The appropriate
counts due to the point sources were included (as fixed models
normalized to the results of the spatial analysis) while carrying out
the fitting of the extended source spectrum as described below.  We
have extracted a spectrum for the cluster candidate using photons
detected within a $30''$ radius of $\alpha = 22^h 15^m 58\fs5,
\delta -17\arcdeg 38\arcmin 2\fs5$.  We extracted MOS1, MOS2 and pn
spectra from each XMM observation with an exposure time $> 10$ ks made
towards the cluster.  We generated a combined spectrum for each camera
and these are presented in the left panel of
Figure~\ref{xspec}. Spectral fitting was carried out using XPSEC
v11.3.1s \citep{xspec} using the MEKAL models \citep{mekal} modified
with an interstellar absorption
\citep{wabs} appropriate for the Galactic column density at that
location. A local background correction was made using a spectrum
extracted from a $30''$ radius region at the same off-axis angle
($9'$) as the cluster; this avoids having to carry out an energy
dependent vignetting correction of the background.  To find the best
fit spectrum, we limited the fit to the energy range $0.5-7$ keV and
grouped the data so that there were a minimum of 20 counts per
bin. The number of background corrected counts in that energy range
(summed across all three cameras) was 1100 in the $30\arcsec$
measurement aperture, after correction for the point sources. We fixed
the hydrogen column density to the
\citet{NH} value, the metal abundance to $Z=0.3$ times the Solar value
and, initially, the redshift to the Keck spectroscopic value,
$z=1.45$.  After accounting for all point sources including the
possible central source, the best fit temperature of the extended
emission was found to be $T_x = 6.5^{+2.6}_{-1.8}$ keV (90\%
confidence limits).  If there is no central point source, then the
best fit temperature of the ICM would be $T_x = 7.4^{+2.7}_{-1.8}$
keV. The best fit model, and its residuals, are shown in
Figure~\ref{xspec}. Using this spectrum, we have determined the
cluster flux to be $1.22^{+0.21}_{-0.17} \times 10^{-14}$ erg s$^{-1}$
cm$^{-2}$ [0.5 -2.0 keV], extrapolated to a region 2 Mpc in radius using the best 
fit $\beta$ model.
For the same 2 Mpc aperture, the luminosity of the ICM is
$1.15^{+0.20}_{-0.16} \times 10^{44}$ erg s$^{-1}$ in the [$0.5 -
2.0$] keV band, and $4.39^{0.76}_{-0.61} \times 10^{44}$ erg s$^{-1}$
in a bolometric band [$0.05-100$ keV].

\subsection{Optical Spectroscopy}

\begin{figure}[h]
\epsscale{0.9}
\plotone{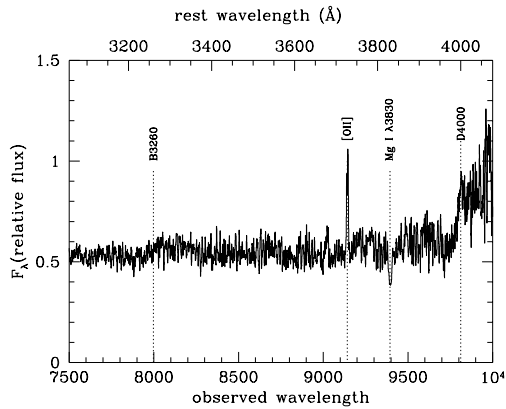}

\caption{
An optical spectrum of cluster member 14339 with $z=1.45$.  The
identified spectral features are marked.  The spectrum has been
smoothed by a 6.5\AA\ boxcar. }

\label{ospec}
\end{figure}

Visual inspection of the reduced spectra from both DEIMOS masks
yielded redshifts for objects.  The 6 galaxies listed in
Table~\ref{members} lie within 30 arcsec of the X-ray centroid and have
concurrent redshifts $1.445 < z_{spec} < 1.455$.  In 3 of the 6
spectra, the [O II]$\lambda~3727$ doublet is detected
and clearly split.  The other three objects have spectra which show
features from the redshifted 4000\AA\ and 2900\AA\ breaks.  The
spectrum of one of the cluster members, ID~14339, is shown in
Figure~\ref{ospec}.  In addition to these 6 galaxies we consider to be
members of the cluster, three more galaxies were spectroscopically
identified with redshifts at $1.44 < z < 1.46$.  However, they lie
approximately $5-7$ arcmin from the X-ray centroid, so are not
(yet) clearly members of the cluster.

\subsection{Optical and Infrared Photometry}

The color--magnitude diagram for all objects in a 3\arcmin~area at the
cluster is shown in Figure~\ref{cmd}.  The objects within 30 arcsec of
the X-ray centroid are circled in red. The 6 galaxies with
concurrent spectroscopic redshifts $1.445 < z < 1.455$ are shown by
the solid red points.  The scatter seen in the colors of the member
galaxies is dominated by the photometric uncertainties in the
photometry so is not indicative of the intrinsic scatter in the red
sequence in this cluster.  Also plotted in Figure~\ref{cmd} are two
estimates of the expected color of L$^*$ early--type galaxies. Using a
simple passive evolution model calculated from the GISSEL of
\citet{BC} and our assumed cosmology, we calculated the expected
colors for a single 100 Myr burst stellar population of solar
metallicity stars formed at $z_f = 3.0$ and $z_f = 2.0$, with slopes 
assumed to be the same as those in Coma for a similar rest frame color.  As can be seen in
Figure~\ref{cmd}, the colors of the member galaxies are consistent
with those exected for early-type galaxies whose stars formed at $2 < z_f <
3.0$.

\section{Discussion}

Analyses of our combined X-ray imaging spectroscopy, optical
spectroscopy, and optical-IR imaging provide strong evidence that
XMMXCS~J2215.9-1738 is a massive galaxy cluster at $z=1.45$.
At least 10 galaxy clusters with $z>1.4$ and $T_{x}>6$ keV are 
expected to be present in our current XCS area of 168.2 deg$^{2}$, if 
the universe is assumed to be flat with $\Omega_{m}=0.238$ and 
$\sigma_{8}=0.744$ (WMAP - Spergel et al. 2006).  This estimate requires the relation 
between cluster mass and X-ray temperature to be normalized so as to 
reproduce the local abundance of X-ray clusters (HIFLUGCS - Reiprich \& 
B\"{o}hringer 2002). However, it should be noted that these 
assumptions require that a local 5 keV cluster would need to 
have a mass, within an overdensity of 500 with respect to the critical 
density, of around $2\times10^{14}\,h^{-1}\,M_{\sun}$ \citep{pier,viana}, 
about 2/3 smaller than is implied by the most recent 
observational data \citep{app,av05}.

With the combination of the serendipitous searches for clusters in the
{\it XMM} archival data \citep{xcs,mullis}, and the on-going wide-area
surveys such as UKIDSS (Lawrence et al.\ in
preparation) and the RCS \citep{RCS} the
time finally is ripe for the identification of large samples of $z >
1$ clusters.  The construction of such samples will pave the way
towards a better understanding of the origin of galaxy clusters and
their constituent galaxy populations.

\acknowledgments

The authors thank the referee for suggestions which improved the 
final version of this paper. 
This work is based on data obtained by XMM-Newton, an ESA science
mission funded by contributions from ESA member states and from NASA.
We acknowledge financial support from; the NASA-LTSA program, the RAS
Hosie Bequest, Liverpool John Moores University, the Institute for
Astronomy at the University of Edinburgh, the XMM \& Chandra guest
observer programmes, Carnegie Mellon University, the NSF and PPARC.
This research made use of the NASA/IPAC Extragalactic Database, the
SIMBAD facility at CDS, the NASA/GSFC supported XSPEC software, and
the ESO Imaging Survey.  The W.\ M.\ Keck Observatory is a scientific
partnership between the University of California and the California
Institute of Technology, made possible by a generous gift of the W.\
M.\ Keck Foundation. The authors wish to recognize and acknowledge the
very significant cultural role and reverence that the summit of Mauna
Kea has always had within the indigenous Hawaiian community; we are
fortunate to have the opportunity to conduct observations from this
mountain.  The analysis pipeline used to reduce the DEIMOS data was
developed at UC-Berkeley with support from NSF grant AST-0071048.  We thank 
the referee for helping to improve the final version of this paper.  The
work by SAS at LLNL was performed under the auspices of the
U.S. Department of Energy under Contract No.\ W-7405-ENG-48. The work
of D.S.\ was carried out at the JPL, California
Institute of Technology, under contract with NASA.

\begin{deluxetable}{ccccll}
\small
\tablenum{1}
\tablecaption{Summary of Spectroscopic Cluster Members}
\tablehead{
\colhead{ID} & 
\colhead{R.A.\tablenotemark{a}} & 
\colhead{Dec.\tablenotemark{a}} & 
\colhead{$K_s$} &
\colhead{$I-K_s$} & 
\colhead{spec-z} 
}
\startdata
14651 & 22:15:58.364 & $-$17:37:37.5 & $>$19.9 & $<$4.0  & 1.4526 \\
14478 & 22:15:57.233 & $-$17:37:53.1 & 19.58 & 4.20 & 1.4537 \\
14378 & 22:15:58.879 & $-$17:37:59.6 & 19.49 & 4.00 & 1.451 \\
14289 & 22:15:58.488 & $-$17:38:10.4 & 19.04 & 4.49 & 1.453 \\
14339 & 22:15:59.057 & $-$17:38:02.0 & 19.18 & 4.16 & 1.4467 \\
14389 & 22:15:58.479 & $-$17:37:58.6 & 19.07 & 4.17 & 1.452 \\

\enddata
\tablenotetext{a}{Coordinates are J2000.}
\label{members}
\end{deluxetable}


\begin{thebibliography}{}

\bibitem[Arnaud(1996)]{xspec}Arnaud, K. in Astronomical Data Analysis Software and Systems V, 1996, 
A.S.P. Conf. Ser., 101,  G. Jacoby \& J. Barnes, eds.,  17

\bibitem[Arnaud, Pointecouteau \& Pratt(2005)]{app} Arnaud, M., Pointecouteau, E., Pratt, G. W., 2005, \aa, 441, 893

\bibitem[Barkhouse {et al.}(2006)]{champ}Barkhouse, W.\ et al.\ 2006, \apj, in press

%\bibitem[Bertin \& Arnout(2001)]{sex}Bertin \& Arnout, \aa, 999, 999

\bibitem[Bremer {et al.}(2006)]{xmmlss}Bremer, W.\ et al.\ 2006, \mnras, in press

\bibitem[Bruzual \& Charlot(2003)]{BC} Bruzual, G. \& Charlot, S.\ 2003, \mnras, 344, 1000

\bibitem[Carlstrom {et al.}(2002)]{sz} Carlstrom, J.~E., Holder, G.~P., \& Reese, E.~D.\ 2002, \araa, 40, 643

\bibitem[Dickey \& Lockman(1990)]{NH}Dickey, J.M. \& Lockman, F.J., 1990, ARA\&A,  28, 215

\bibitem[Dietrich {et~al.}(2005)]{eis} Dietrich, J.P. et al.\ 2005, astro-ph/0510223

\bibitem[Faber {et~al.}(2003)]{deimos} Faber, S.\ et al.\ 2003, SPIE, 4841, 1657

\bibitem[Gladders \& Yee(2005)]{RCS}Gladders, M.\ \& Yee, H.\ 2005, \apjs, 157, 1

\bibitem[McCammon  \& Sanders(1990)]{wabs}McCammon, D. \& Sanders, D.T.\ 1990, ARA\&A, 28, 657

\bibitem[Mewe {et al.}(1986)]{mekal}Mewe, R., Lemen, J.R. \& van der Oord, G.H.J., 1986, A \& AS, 62, 197

\bibitem[Mullis {et al.}(2005)]{mullis}Mullis, C. et al.\ 2005, \apj, 623, L85

\bibitem[Pierpaoli {et al.}(2003)]{pier} Pierpaoli, E., Borgani, S., Scott, D., \& White, M., 2003, MNRAS, 342, 163

\bibitem[Reiprich \& B\"{o}hringer(2002)]{hiflugcs} Reiprich, T. H.\ \& B\"{o}hringer, H., 2002, \apj, 567, 716

\bibitem[Romer {et al.}(2001)]{xcs}Romer, K., Viana, P.T.P, Liddle, A.R., \& Mann, R.G.\ 2001, \apj, 547, 594 

\bibitem[Rosati {et al.}(2004)]{1252}Rosati, P. et al.\ 2004, \aj, 127, 230

\bibitem[Spergel {et al. }(2006)]{wmap3} Spergel, D.N. et al., 2006, submitted to ApJ, astro-ph/0603449

\bibitem[Stanford {et al.}(2001)]{lynx}Stanford, S.A., et al.\ 2001, \apj, 552, L504

\bibitem[Stanford {et al.}(2002)]{0910}Stanford, S.A., Holden, B.,Rosati, P., Eisenhardt, P., Stern, D., Squires, G., \& Spinrad, H.\ 2002, \aj, 123, 619

\bibitem[Stanford {et al.}(2005)]{bootes}Stanford, S.A.\ et al.\ 2005, \apj, 634, L129

\bibitem[Viana {et al. }(2003)]{viana} Viana, P. T. P., Kay, S. T., Liddle, A. R., Muanwong, O., Thomas, P. A., 2003, MNRAS, 346, 319

\bibitem[Vikhlinin {et al. }(2006)]{av05} Vikhlinin, A., Kravtsov, A., Forman, W., Jones, C., Markevitch, M., Murray, S. S., Van Speybroeck, L., 2006, \apj, 640, 661  

\end{thebibliography}
\end{document}